# Generalized Method of Moments and Percentile Method: Estimating parameters of the Novel Median Based Unit Weibull Distribution

Iman M. Attia *

Imanattiathesis1972@gmail.com ,imanattia1972@gmail.com

*Department of Mathematical Statistics, Faculty of Graduate Studies for Statistical Research, Cairo University, Egypt

*Abstract*: The Median Based Unit Weibull is a new 2 parameter unit Weibull distribution defined on the unit interval (0,1). Estimation of the parameters using MLE encountered some problems like large variance. Using generalized method of moments (GMMs) and percentile method may ameliorate this condition. This paper introduces GMMs and the percentile methods for estimating the parameters of the new distribution with illustrative real data analysis.

# Introduction

Waloddi Weibull(1951) was the first to introduce the Weibull distribution. It is one of the famous distributions used to model life data and reliability. It can describe the increasing failure rate cases as well as the decreasing failure rate cases. The exponential distribution is a special case of it, when the shape parameter is one. Rayleigh distribution is another special case of it, when the shape parameter is 2. It can also describe and explain the life expectancy of the elements entailed in the fatigue derived failure and can also evaluate the electron tube reliability and load handling machines. It is used in many fields like medicine, physics, engineering, biology, and quality control. As the distribution does not represent a bathtub or unimodal shapes, this enforces many researchers to generalize and transform this distribution in the recent decades. To mention some of these researchers:, (Singla et al., 2012) elucidated beta generalized weibull, (Khan et al., 2017) described in details the transmuted weibull, (Xie et al., 2002) explored the modified weibull, (Lee et al., 2007) clearly explained the beta weibull, (Cordeiro et al., 2010) demonstrated Kumaraswamy weibull, (Silva et al., 2010) expounded beta modified weibull, (Mudholkar & Srivastava, 1993) expatiated the exponentiated weibull, (Zhang & Xie, 2011) interpreted truncated weibull, (Khan & King, 2013) explicated transmuted modified weibull, and (Marshall & Olkin, 1997) handled the extended weibull.

The generalized method of moments has been developed by many authors like (Benson, M.A., 1968) and the United States Water Resources Council (US Water Resources Council, 1967; US Water Resources Council, 1981) , (Bob~e, B., 1975) , (Hoshi, K. and Burges,

S.J., 1981) , (Rao, D.V, 1980) , (Ashkar, F. and Bob~e, a., 1986a; Ashkar, F. and Bob~e, B., 1986B) , (Bob~e, B. and Ashkar, F, 1986) used method of SAM. All these authors applied the most famous four methods of generalized method of moments to the Log Pearson type 3 (LP) distribution. But these methods can be applied to any given distribution and not specific to the LP distribution. As shown by (Shenton, L.R. and Bowman, K.O, 1977) , the maximum likelihood estimators are moment estimators in case of LP distribution. (Ashkar et al., 1988) introduced GMMs for estimation of the parameters of generalized gamma distribution.

Estimation using the method of percentiles has been used for a long time. It has been used in estimation by many authors and has been found to be better or equally efficient to MLE and least squares (Bhatti et al., 2018; Dubey, 1967; Marks, 2005; Wang & Keats, 1995).

In this paper, the author will discuss the GMMs and the percentile method to estimate the parameters of the new distribution MBUW. The author will also discuss application of these methods on real data.

The paper is arranged into 4 sections. In section 1, the author will explain the methodology of estimating the parameters using GMMs. In section 2, elaboration of the percentile method of estimation applied to the MBUW distribution. In section 3, analysis of some real data sets to illustrate to what extent the distribution can fit the data. In section 4, conclusion and future work will be declared.

## Section 1

## Methodology:

## Generalized Method of Moments (GMMs):

The BMUW distribution has been discussed in previous work by the author as regards properties and some methods of estimation with applications on real data analysis. The new distribution has the following pdf, cdf and quantilel function respectively,

$$f(y) = \frac{6}{\alpha^\beta}\left[1 - y^{\frac{1}{\alpha^\beta}}\right] y^{\left(\frac{2}{\alpha^\beta}-1\right)}, \quad 0 < y < 1, \quad \alpha > 0, \beta > 0 \quad (1)$$

$$F(y) = 3y^{\frac{2}{\alpha^\beta}} - 2y^{\frac{3}{\alpha^\beta}}, \quad 0 < y < 1, \quad \alpha > 0, \beta > 0 \quad (2)$$

$$u = F(y) = 3y^{\frac{2}{\alpha^\beta}} - 2y^{\frac{3}{\alpha^\beta}} = -2\left(y^{\frac{1}{\alpha^\beta}}\right)^3 + 3\left(y^{\frac{1}{\alpha^\beta}}\right)^2 \quad (3)$$

The inverse of the CDF (quantile) is used to obtain y, the real root of this 3rd polynomial function is :

$$y = F^{-1}(y) = \left\{-.5\left(\cos\left[\frac{\cos^{-1}(1-2u)}{3}\right] - \sqrt{3}\sin\left[\frac{\cos^{-1}(1-2u)}{3}\right]\right) + .5\right\}^{\alpha^\beta} \quad (4)$$

GMMs uses the raw moments of the distribution. Equating the sample moments with the population moments is the main corner of the method. As the MBUW has 2 parameters, the first and second moments will be used to form a system of equations solved numerically to find the parameters.

Steps of algorithm:

1. The non-central sample moments of order r for a given sample $y_1$, $y_2$,...., $y_N$ are defined as :

$$m'_r(y) = \frac{1}{N} \sum_{i=1}^{r} y_i^r \quad \ldots\ldots\ldots\ldots\ldots (5)$$

2. For this distribution, the non-central population moment has the following formula:

$$\mu'_r = E(y^r) = \frac{6}{(2 + r\alpha^\beta)(3 + r\alpha^\beta)} \quad \ldots\ldots\ldots\ldots (6)$$

$$\mu'_1 = E(y) = \frac{6}{(2 + \alpha^\beta)(3 + \alpha^\beta)} \quad \ldots\ldots (7)$$

$$\mu'_2 = E(y^2) = \frac{6}{(2 + 2\alpha^\beta)(3 + 2\alpha^\beta)} \quad \ldots\ldots (8)$$

3. Equate the sample moment and the population moment for the first and second order.

$$m'_r(y) = \mu'_r \quad \ldots\ldots\ldots (9)$$

$m'_1(y) = \mu'_1$, this leads to the following equation:

$$m'_1(y) = \mu'_1 = \frac{1}{N}\sum_{i=1}^{r} y = \bar{y} = \frac{6}{(2+\alpha^\beta)(3+\alpha^\beta)} \quad \ldots \ldots (10)$$

$m'_2(y) = \mu'_2$, this leads to the following equation:

$$m'_2(y) = \mu'_2 = \frac{1}{N}\sum_{i=1}^{r} y^2 = \frac{6}{(2+2\alpha^\beta)(3+2\alpha^\beta)} \quad \ldots \ldots \ldots (11)$$

4- Differentiate equation 10 and 11 with respect to both parameters and solve the system of equations using Levenberg-Marquardt algorithm (LM).

$$\frac{\partial}{\partial \alpha}\mu'_1 = \bar{y}\left(5\beta\alpha^{\beta-1} + 2\beta\alpha^{2\beta-1}\right) - 6 = 0 \quad \ldots \ldots \ldots (12)$$

$$\frac{\partial}{\partial \beta}\mu'_1 = \bar{y}\left(5\alpha^\beta \ln(\alpha) + 2\alpha^{2\beta}\ln(\alpha)\right) - 6 = 0 \quad \ldots \ldots \ldots (13)$$

$$\frac{\partial}{\partial \alpha}\mu'_2 = m'_2\left(10\beta\alpha^{\beta-1} + 8\beta\alpha^{2\beta-1}\right) - 6 = 0 \ldots \ldots \ldots (14)$$

$$\frac{\partial}{\partial \beta}\mu'_2 = m'_2\left(10\alpha^\beta \ln(\alpha) + 4\alpha^{2\beta}\ln(\alpha)\right) - 6 = 0 \quad \ldots \ldots \ldots (15)$$

The objective function to be minimized is

$$E(y^r) = \frac{6}{(2+r\alpha^\beta)(3+r\alpha^\beta)}$$

$$\theta^{(a+1)} = \theta^{(a)} + \left[J'(\theta^{(a)})J(\theta^{(a)}) + \lambda^{(a)}I^{(a)}\right]^{-1}\left[J'(\theta^{(a)})\right]\left(y - f(\theta^{(a)})\right) \ldots (16)$$

Where:

The Jacobian matrix is $\begin{bmatrix} equation12 & equation14 \\ equation13 & equation15 \end{bmatrix}$

$$\left(y - f(\theta^{(a)})\right) = \begin{bmatrix} m'_1 - \dfrac{6}{(2+\alpha^\beta)(3+\alpha^\beta)} \\ m'_2 - \dfrac{6}{(2+2\alpha^\beta)(3+2\alpha^\beta)} \end{bmatrix}$$

Where the parameters used in the first iteration are the initial guess, then they are updated according to the sum of squares of errors.

LM algorithm is an iterative algorithm.

$J'(\theta^{(a)})$ : this is the Jacobian function which is the first derivative of the objective function evaluated at the initial guess $\theta^{(a)}$.

$\lambda^{(a)}$ is a damping factor that adjust the step size in each iteration direction, the starting value usually is 0.001 and according to the sum square of errors (SSE) in each iteration this damping factor is adjusted:

$SSE^{(a+1)} \geq SSE^{(a)}$   so update:   $\lambda_{updated} = 10 * \lambda_{old}$

$SSE^{(a+1)} < SSE^{(a)}$   so update:   $\lambda_{updated} = \dfrac{1}{10} * \lambda_{old}$

$f(\theta^{(a)})$ : is the objective function (non-central population moment, first and second order) evaluated at the initial guess.

$y$ : is the non-central sample moments, the first and second order.

Steps of algorithm:

1- Start with the initial guess of parameters (alpha and beta).

2- Substitute these values in the objective function and the Jacobian.

3- Choose the damping factor, say lambda=0.001

4- Substitute in equation (LM equation (16)) to get the new coefficients.

5- Calculate the SSE at these parameters and compare this SSE value with the previous one when using initial parameters to adjust for the damping factor.

6- Update the damping factor accordingly as previously explained.

7- Start new iteration with the new parameters and the new updated damping factor, i.e , apply the previous steps many times till convergence is achieved or a pre-specified number of iterations is accomplished.

The value of this quantity: $[J'(\theta^{(a)})J(\theta^{(a)}) + \lambda^{(a)}I^{(a)}]^{-1}$ can be considered a good approximation to the variance – covariance matrix of the estimated coefficients. Standard errors for the estimated coefficients are the square root of the diagonal of the elements in this matrix.

## Section 2

### The Percentile Method

The percentile method used in this paper is applied to the 25$^{th}$ and 75$^{th}$ percentiles. The MBUW has the quantile or the percentile function

$$y = \left\{-.5\left(\cos\left[\frac{\cos^{-1}(1-2u)}{3}\right] - \sqrt{3}\sin\left[\frac{\cos^{-1}(1-2u)}{3}\right]\right) + .5\right\}^{\alpha^{\beta}}$$

$$c = -.5\left(\cos\left[\frac{\cos^{-1}(1-2u)}{3}\right] - \sqrt{3}\sin\left[\frac{\cos^{-1}(1-2u)}{3}\right]\right) + .5, 0 < u < 1 \quad (17)$$

$$y_{25} = \left\{-.5\left(\cos\left[\frac{\cos^{-1}(1-2*0.25)}{3}\right] - \sqrt{3}\sin\left[\frac{\cos^{-1}(1-2*0.25)}{3}\right]\right) + .5\right\}^{\alpha^{\beta}} \quad (18)$$

$$y_{75} = \left\{-.5\left(\cos\left[\frac{\cos^{-1}(1-2*.75)}{3}\right] - \sqrt{3}\sin\left[\frac{\cos^{-1}(1-2*.75)}{3}\right]\right) + .5\right\}^{\alpha^{\beta}} \quad (19)$$

$$y_{25} = c_{25}^{\alpha^{\beta}}, \quad y_{75} = c_{25}^{\alpha^{\beta}} \quad (20)$$

The objective function to be minimized in LM algorithm is the quantile function. Differentiation of this objective function with respect to both parameters gives these equations:

$$\frac{\partial}{\partial \alpha} y_{25} = c_{25}^{\alpha^{\beta}} [\ln(c_{25})] \beta \alpha^{\beta-1} \quad (21)$$

$$\frac{\partial}{\partial \beta} y_{25} = c_{25}^{\alpha^{\beta}} [\ln(c_{25})] (\alpha^{\beta})[\ln(\alpha)] \quad (22)$$

$$\frac{\partial}{\partial \alpha} y_{75} = c_{75}^{\alpha^{\beta}} [\ln(c_{75})] \beta \alpha^{\beta-1} \quad (23)$$

$$\frac{\partial}{\partial \beta} y_{75} = c_{75}^{\alpha^{\beta}} [\ln(c_{75})] (\alpha^{\beta})[\ln(\alpha)] \quad (24)$$

The Jacobian matrix is $\begin{bmatrix} equation21 & equation23 \\ equation22 & equation24 \end{bmatrix}$

$$\left(y - f(\theta^{(a)})\right) = \begin{bmatrix} y_{25} - c_{25}^{\alpha^{\beta}} \\ y_{75} - c_{75}^{\alpha^{\beta}} \end{bmatrix}$$

Then use LM algorithm as previously explained.

Any percentiles can be used, provided 2 consecutive percentiles. The most common to use are $25^{th}$ and $75^{th}$ percentiles.

# Section 3

**Some real data analysis:**

The database, OECD, is used. Some variables are analyzed to discover what distributions fit the data better. The data is available at:

https://stats.oecd.org/index.aspx?DataSetCode=BLI

It was used by the author in other works .(Iman M.Attia, 2024)

*First data* : (Dwelling Without Basic Facilities)

These observations measure the percentage of homes in the involved countries that lack essential utilities like indoor plumbing, central heating, clean drinking water supplies.

| 0.008 | 0.007 | 0.002 | 0.094 | 0.123 | 0.023 | 0.005 | 0.005 | 0.057 | 0.004 |
|---|---|---|---|---|---|---|---|---|---|
| 0.005 | 0.001 | 0.004 | 0.035 | 0.002 | 0.006 | 0.064 | 0.025 | 0.112 | 0.118 |
| 0.001 | 0.259 | 0.001 | 0.023 | 0.009 | 0.015 | 0.002 | 0.003 | 0.049 | 0.005 |
| 0.001 | 0.03 | 0.067 | 0.138 | 0.359 | | | | | |

*Second data* : (Quality of Support Network)

This data set explores how much the person can rely on sources of support like family, friends, or community members in time of need and disparate. It is represented as percentage of persons who had found social support in times of crises.

| 0.92 | 0.93 | 0.88 | 0.80 | 0.82 | 0.96 | 0.95 | 0.96 | 0.94 | 0.90 |
|------|------|------|------|------|------|------|------|------|------|
| 0.78 | 0.98 | 0.89 | 0.92 | 0.91 | 0.77 | 0.94 | 0.95 | 0.96 | 0.85 |

### *Third data* : ( Voter Turnout )

This data set evaluates the percentage of capable and qualified persons for casting a vote in election reflecting the democracy in the country.

| 0.92 | 0.76 | 0.88 | 0.68 | 0.47 | 0.53 | 0.66 | 0.62 | 0.85 | 0.64 |
|------|------|------|------|------|------|------|------|------|------|
| 0.69 | 0.75 | 0.79 | 0.58 | 0.70 | 0.81 | 0.63 | 0.67 | 0.73 | 0.53 |
| 0.77 | 0.55 | 0.57 | 0.90 | 0.63 | 0.79 | 0.82 | 0.78 | 0.68 | 0.49 |
| 0.66 | 0.53 | 0.72 | 0.87 | 0.45 | 0.86 | 0.68 | 0.65 |      |      |

### *Fourt data* :  ( Flood Data)

These are 20 observations for the maximum flood level in Susquehanna River at Harrisburg, Penssylvania (Dumonceaux & Antle, 1973) .

| 0.26 | 0.27 | 0.3  | 0.32 | 0.32 | 0.34 | 0.38 | 0.38 | 0.39 | 0.4  |
|------|------|------|------|------|------|------|------|------|------|
| 0.41 | 0.42 | 0.42 | 0.42 | 0.45 | 0.48 | 0.49 | 0.61 | 0.65 | 0.74 |

### *Fifth data* : ( Time between Failures of Secondary Reactor Pumps)(Maya et al., 2024, 1999)(Suprawhardana and Prayoto)

| 0.216  | 0.015  | 0.4082 | 0.0746 | 0.0358 | 0.0199 | 0.0402 | 0.0101 | 0.0605 |
|--------|--------|--------|--------|--------|--------|--------|--------|--------|
| 0.0954 | 0.1359 | 0.0273 | 0.0491 | 0.3465 | 0.007  | 0.656  | 0.106  | 0.0062 |
| 0.4992 | 0.0614 | 0.532  | 0.0347 | 0.1921 |        |        |        |        |

### *Sixth data* : ( to evaluate the factors concerning the unit capacity, data was collected to compare between algorithms like SC 16 and P3 )(Maya et al., 2024, 1999)

| 0.853 | 0.759 | 0.866 | 0.809 | 0.717 | 0.544 | 0.492 | 0.403 | 0.344 |
|-------|-------|-------|-------|-------|-------|-------|-------|-------|
| 0.213 | 0.116 | 0.116 | 0.092 | 0.07  | 0.059 | 0.048 | 0.036 | 0.029 |
| 0.021 | 0.014 | 0.011 | 0.008 | 0.006 |       |       |       |       |

Fitting the MBUW to the above data revealed results with high variance. The method used was the MLE. These data was previously analyzed by the author for fitting the following unit distributions: Beta distribution, Kumaraswamy distribution, Median Based Unit Rayleigh (BMUR) distribution, and Median Based Unit Weibull (MBUW) distribution. The following table shows the results for fitting the MBUW. This was previously discussed in other author's works.(Iman M.Attia, 2024),

These are the pdfs of the distributions used in the analysis of these 6 data sets, in addition to two distributions that were used in analysis of other data sets provided by the author in a previous preprint paper discussing the new distribution (MBUR). (Iman M. Attia, 2024)

1- Beta Distribution:

$$f(y;\alpha,\beta) = \frac{\Gamma(\alpha+\beta)}{\Gamma(\alpha)\Gamma(\beta)} y^{\alpha-1}(1-y)^{\beta-1}, 0 < y < 1, \alpha > 0, \beta > 0$$

2- Kumaraswamy Distribution:

$$f(y;\alpha,\beta) = \alpha\beta y^{\alpha-1}(1-y^\alpha)^{\beta-1}, 0 < y < 1, \quad \alpha > 0, \beta > 0$$

3- Median Based Unit Rayleigh:

$$f(y;\alpha) = \frac{6}{\alpha^2}\left[1 - y^{\frac{1}{\alpha^2}}\right] y^{\left(\frac{2}{\alpha^2}-1\right)}, \quad 0 < y < 1, \quad \alpha > 0$$

Tools of comparison are:

(k) is the number of parameter while (n) is the number of observations.

$$AIC = -2MLL + 2k \quad , where\ k = number\ of\ parameters$$

$$AIC - corrected = -2MLL + \frac{2k}{n-k-1}$$

$$HQIC = 2\log\{\log(n)[k - 2MLL]\}$$

$$BIC = -2MLL + k\log(n)$$

$$KS - test = Sup_n |F_n - F|, \quad F_n = \frac{1}{n}\sum_{i=1}^{n} I_{x_i < x}$$

First, second, third data set:

| | First data set , n=35 | | Second data set n=20 | | Third data set , n=38 | |
|---|---|---|---|---|---|---|
| theta | $\alpha = 6.636$ | | $\alpha = 0.257$ | | $\alpha = 0.5509$ | |
| | $\beta = 0.8726$ | | $\beta = 1.5073$ | | $\beta = 1.2779$ | |
| Var | $3.5 * 10^6$ | $-2.5 * 10^5$ | | | | |
| | $-2.5 * 10^5$ | $1.7 * 10^4$ | | | | |
| SE | 316.227 | | Cannot be estimated | | Cannot be estimated | |
| | 22.039 | | Cannot be estiamted | | Cannot be estiamted | |
| AIC | 152.585 | | 64.079 | | 48.1377 | |
| AIC correc | 152.9600 | | 64.7848 | | 48.4805 | |
| BIC | 155.6957 | | 66.0704 | | 51.4128 | |
| HQIC | 4.2965 | | 3.927 | | 4.0216 | |
| NLL | -74.2925 | | -30.0395 | | -22.0688 | |
| K-S Value | 0.1794 | | 0.1309 | | 0.1364 | |
| $H_0$ | Fail to Reject | | Fail to Reject | | Fail to Reject | |
| P-value | 0.1860 | | 0.8399 | | 0.4401 | |

Fourth, fifth, sixth data set:

| | Fourth data set, n=20 | | Fifth data set n=23 | | Sixth data set, n=23 | |
|---|---|---|---|---|---|---|
| theta | $\alpha = 1.0932$ | | $\alpha = 5.1285$ | | $\alpha = 3.7003$ | |
| | $\beta = 0.9719$ | | $\beta = 0.7113$ | | $\beta = 0.7415$ | |
| Var | $1.66 * 10^4$ | $-1.7 * 10^5$ | $1.02 * 10^7$ | $-8.7 * 10^5$ | $5.2 * 10^6$ | $-7.8 * 10^5$ |
| | $-1.7 * 10^5$ | $1.6 * 10^6$ | $-8.7 * 10^5$ | $7.4 * 10^4$ | $-7.8 * 10^5$ | $1.2 * 10^5$ |
| SE | 28.81 | | 665 | | 475 | |
| | 28.81 | | 56.7 | | 72.2 | |
| AIC | 16.9233 | | 43.862 | | 19.2158 | |
| AIC correc | 17.6292 | | 44.4620 | | 19.8158 | |
| BIC | 18.9148 | | 46.1330 | | 21.4867 | |
| HQIC | 3.4805 | | 3.8543 | | 3.5773 | |
| NLL | -6.4617 | | -19.9310 | | -7.6079 | |
| K-S Value | 0.3202 | | 0.1584 | | 0.1518 | |
| $H_0$ | Fail to Reject | | Fail to Reject | | Fail to Reject | |
| P-value | 0.0253 | | 0.5575 | | 0.4074 | |

As shown from the above analysis, MLE yielded large variance. GMMs are used as previously described in section 3. The initial guess used in LM algorithm was the values obtained from MLE method as shown in the table for each data set. The variances obtained from GMMs method are dramatically small compared to one used by MLE for each data set. The parameter estimation results were assessed by goodness of fit procedures like: KS-test and visualized by qq-plot and the pp-plot.

The following table shows the results of GMMs:

|  | First data set , n=35 | | Second data set n=20 | | Third data set , n=38 | |
|---|---|---|---|---|---|---|
| theta | $\alpha = 7.2434$ | | $\alpha = 0.2381$ | | $\alpha = 0.5474$ | |
|  | $\beta = 0.8744$ | | $\beta = 1.4903$ | | $\beta = 1.281$ | |
| Var | 12.1458 | −0.7633 | 0.0555 | -0.0512 | 0.033 | -0.0292 |
|  | −0.7633 | 0.0617 | -0.0512 | 0.0699 | -0.0292 | 0.0441 |
| SE | 0.586 | | 0.053 | | 0.029 | |
|  | 0.042 | | 0.059 | | 0.034 | |
| SSE | 0.001 | | 0.0003 | | 0.000001 | |
| K-S Value | 0.1804 | | 0.157 | | 0.1409 | |
| $H_0$ | Fail to Reject | | Fail to Reject | | Fail to Reject | |
| P-value | 0.081 | | 0.6515 | | 0.4008 | |
| $m'_1$ | 0.0475 | | 0.9005 | | 0.6911 | |
| $m'_2$ | 0.0081 | | 0.8148 | | 0.4928 | |

| | Fourth data set, n=20 | | Fifth data set n=23 | | Sixth data set, n=23 | |
|---|---|---|---|---|---|---|
| theta | $\alpha = 1.1955$ | | $\alpha = 5.0368$ | | $\alpha = 3.4012$ | |
| | $\beta = 1.056$ | | $\beta = 0.7077$ | | $\beta = 0.7116$ | |
| Var | 0.1468 | −0.1623 | 1.3712 | −0.1323 | 0.2279 | −0.0554 |
| | −0.1623 | 0.2404 | −0.1323 | 0.0248 | −0.0554 | 0.0264 |
| SE | 0.086 | | 0.244 | | 0.099 | |
| | 0.109 | | 0.175 | | 0.034 | |
| SSE | 0.0035 | | 0.00007 | | 0.0106 | |
| K-S Value | 0.2694 | | 0.1682 | | 0.2046 | |
| $H_0$ | Fail to Reject | | Fail to Reject | | Fail to Reject | |
| P-value | 0.0899 | | 0.4822 | | 0.2537 | |
| $m'_1$ | 0.4225 | | 0.1578 | | 0.2881 | |
| $m'_2$ | 0.1932 | | 0.0606 | | 0.1798 | |

The following figures illustrate the ecdf v.s. the theoretical cdf, the pp-plots, and the qq-plots for the MBUW fitted data using the GMMs.

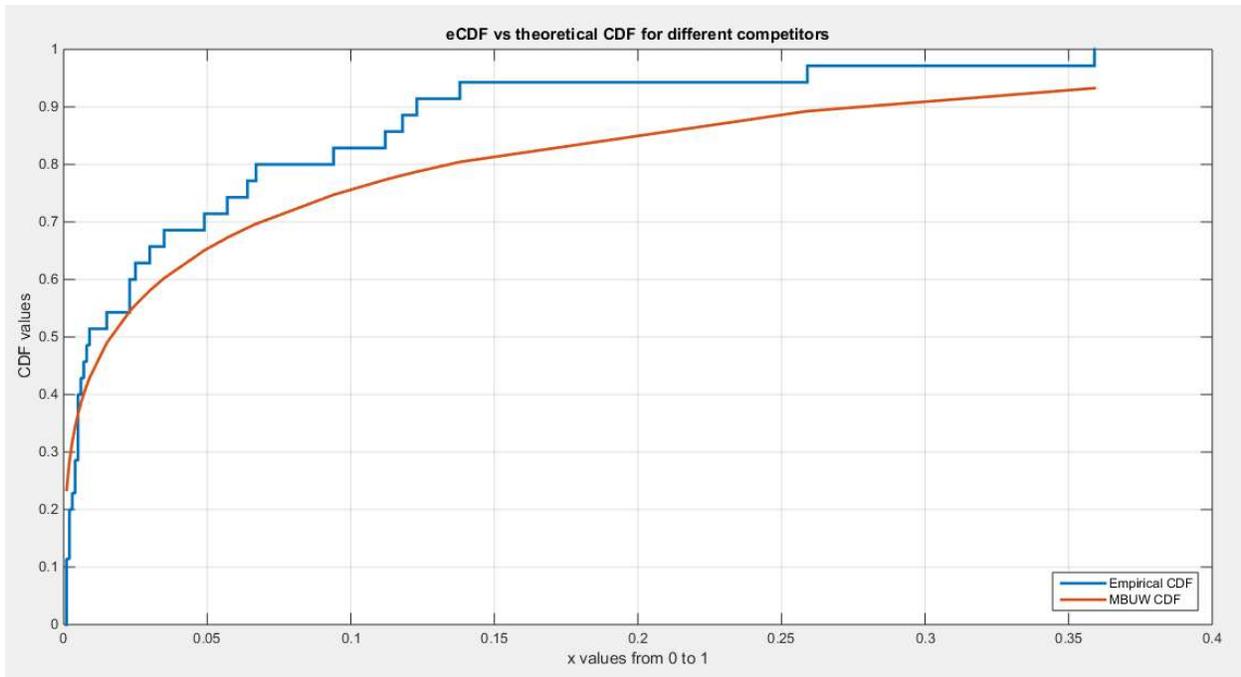

Fig. 1: shows the eCDF vs. theoretical CDF of the BMUW distributions for the 1st data set.

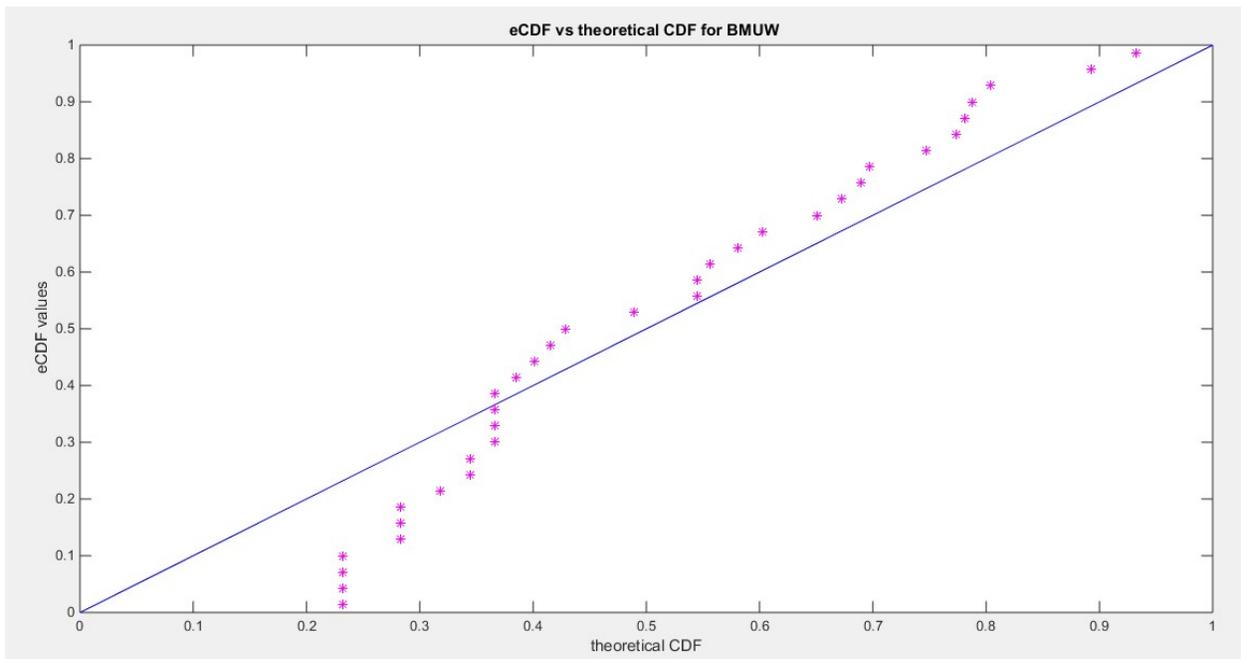

Fig. 2: shows the pp-plot of the BMUW distributions for the 1st data set.

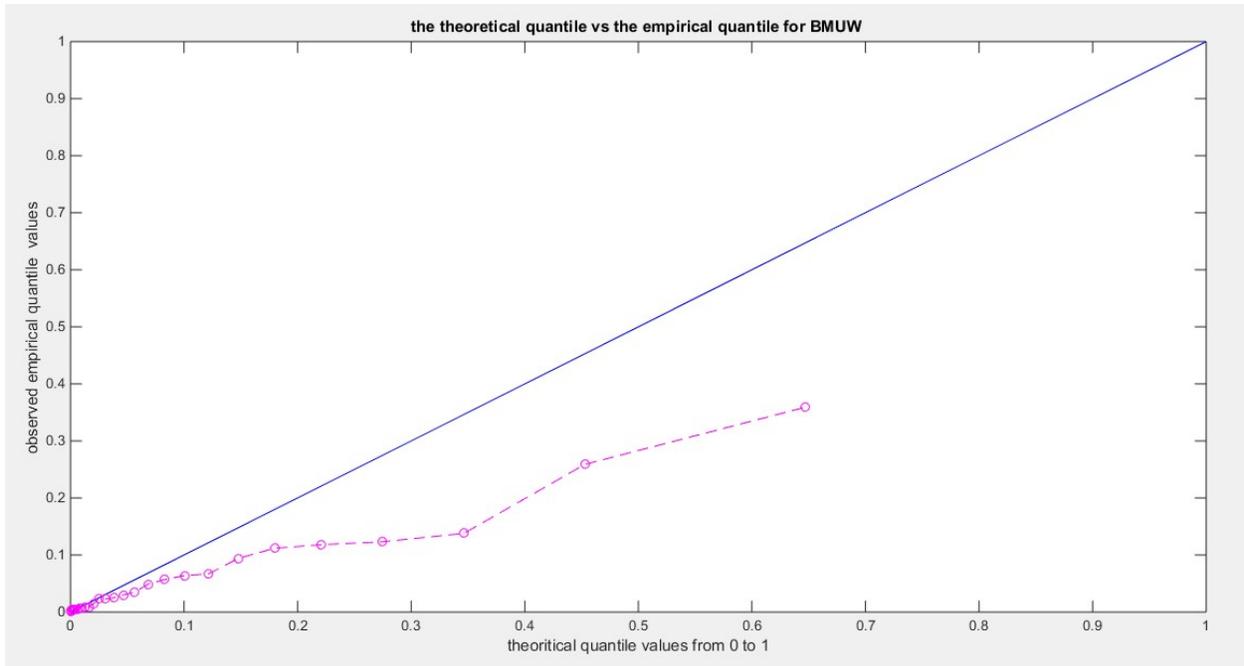

Fig. 3: shows the qq-plot of the BMUW distributions for the 1st data set

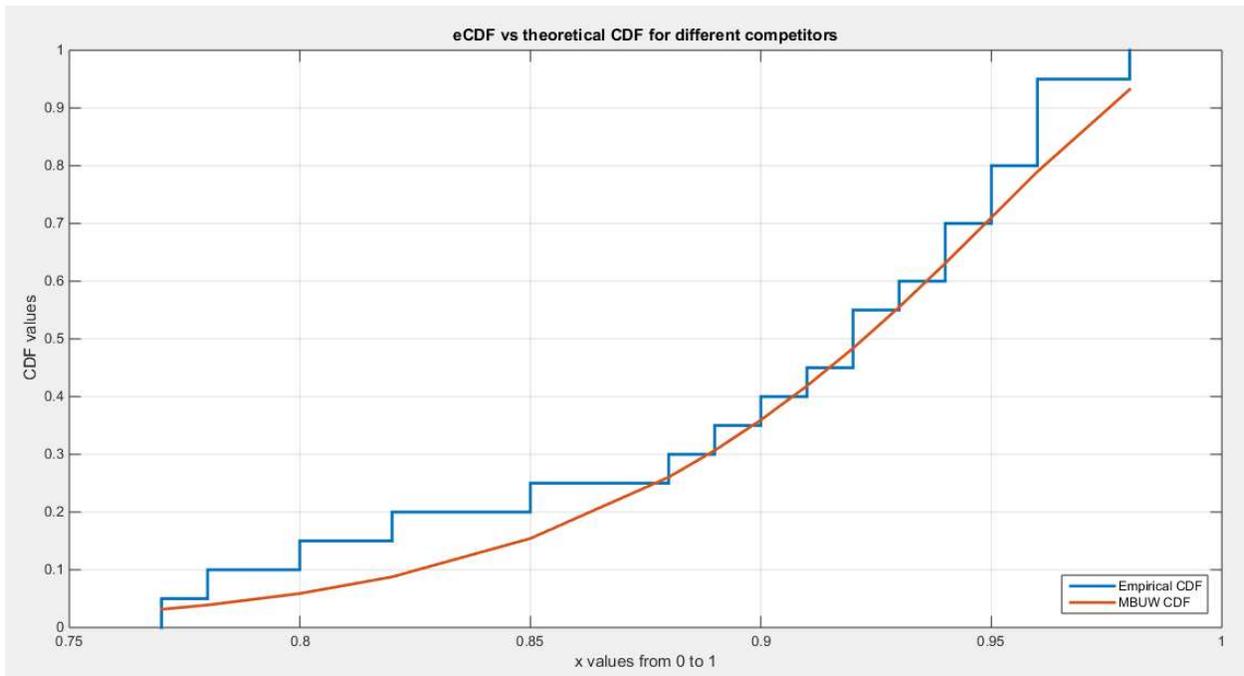

Fig. 4: shows the eCDF vs. theoretical CDF of the BMUW distributions for the 2nd data set.

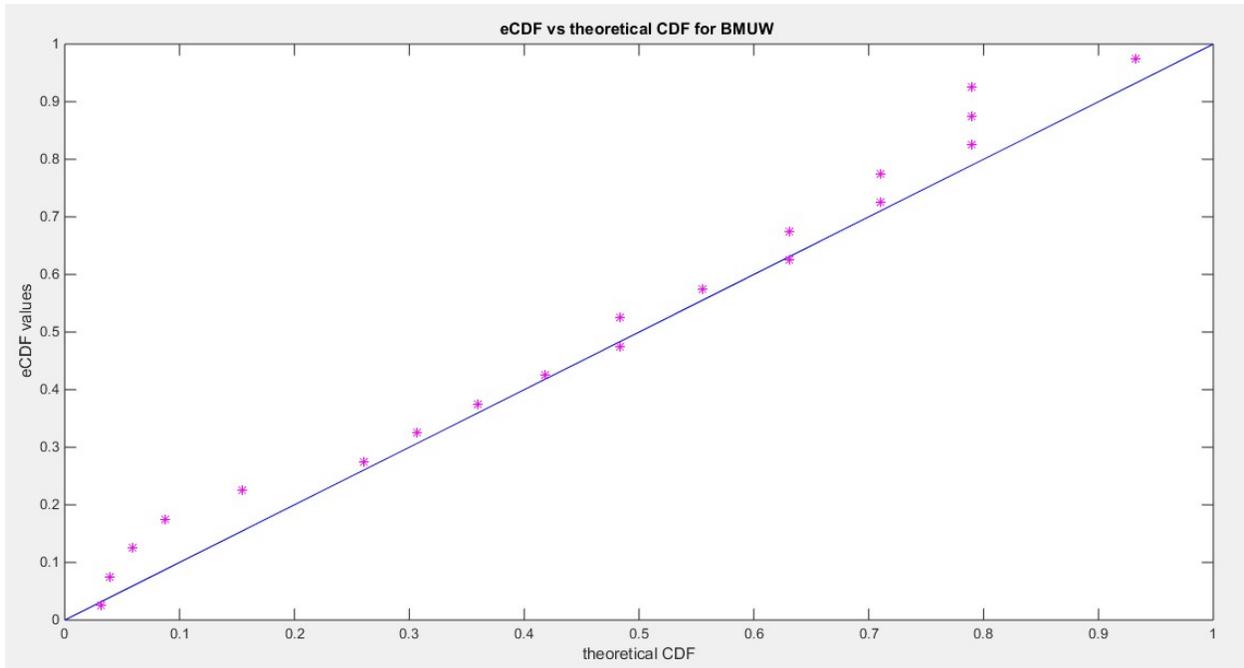

Fig. 5: shows the pp-plot of BMUW distributions for the 2nd data set.

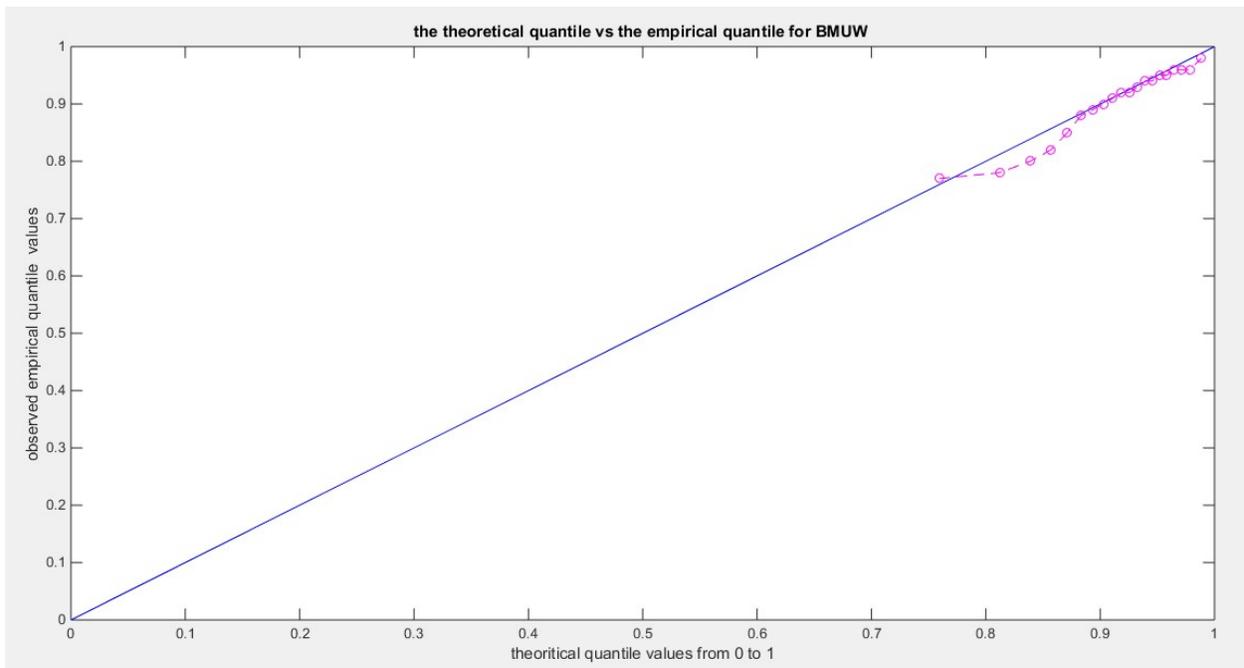

Fig. 6: shows the qq-plot of BMUW distributions for the 2nd data set.

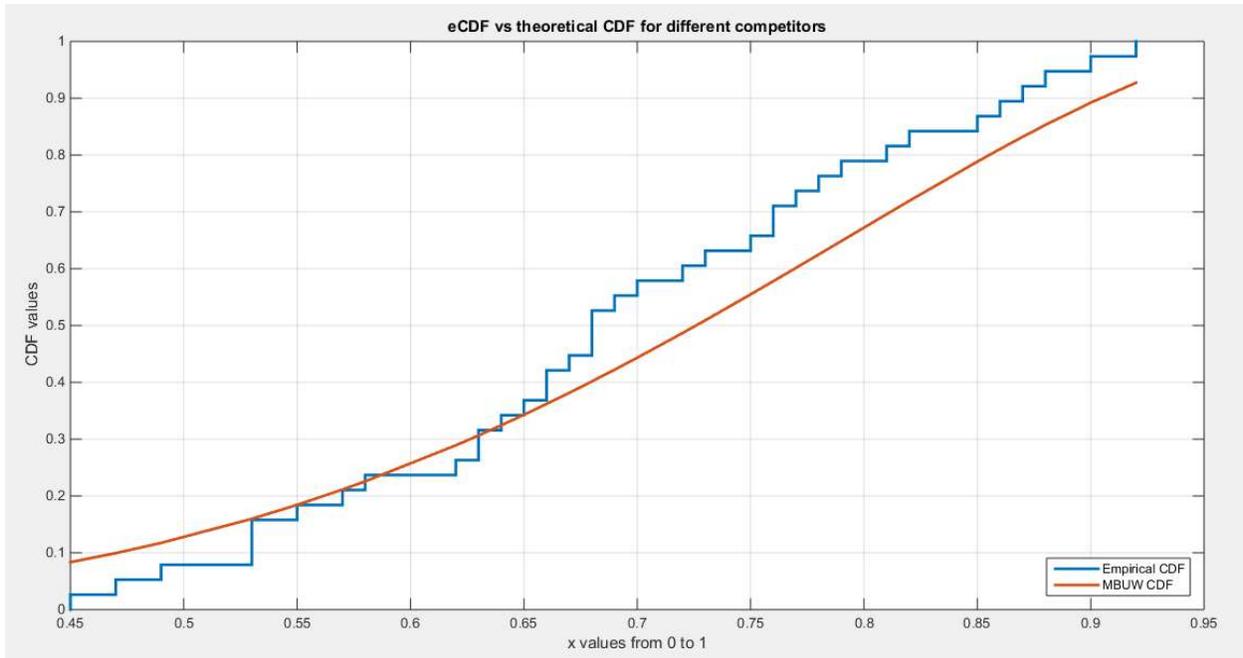

Fig. 7: shows eCDF vs. theoretical CDF of the BMUW distributions for the 3$^{rd}$ data set.

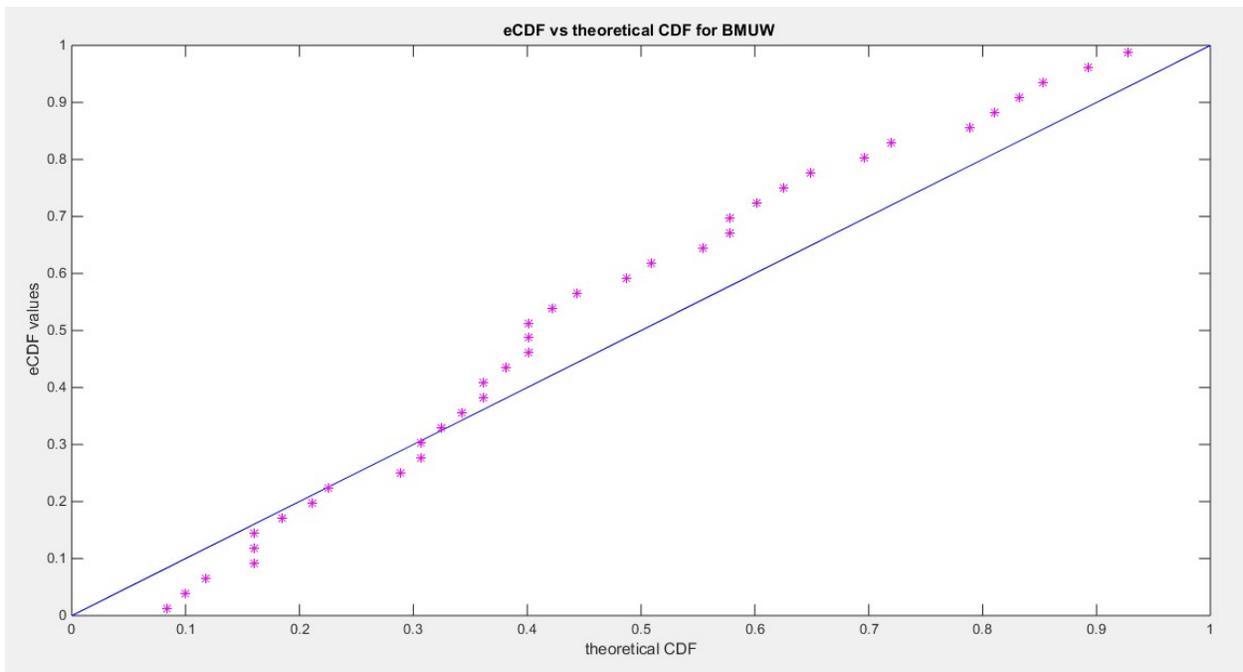

Fig. 8 :shows the pp-plot of BMUW distributions for the 3$^{rd}$ data set.

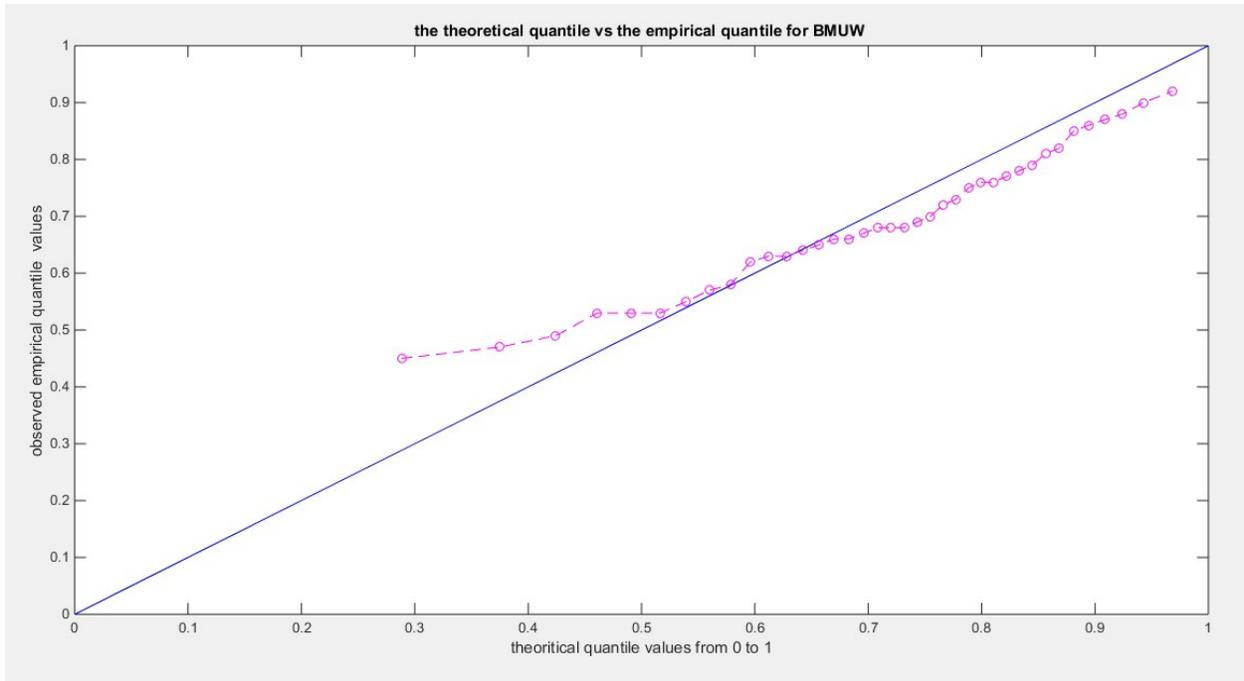

Fig. 9 : shows the qq-plot of BMUW distributions for the 3${}^{rd}$ data set.

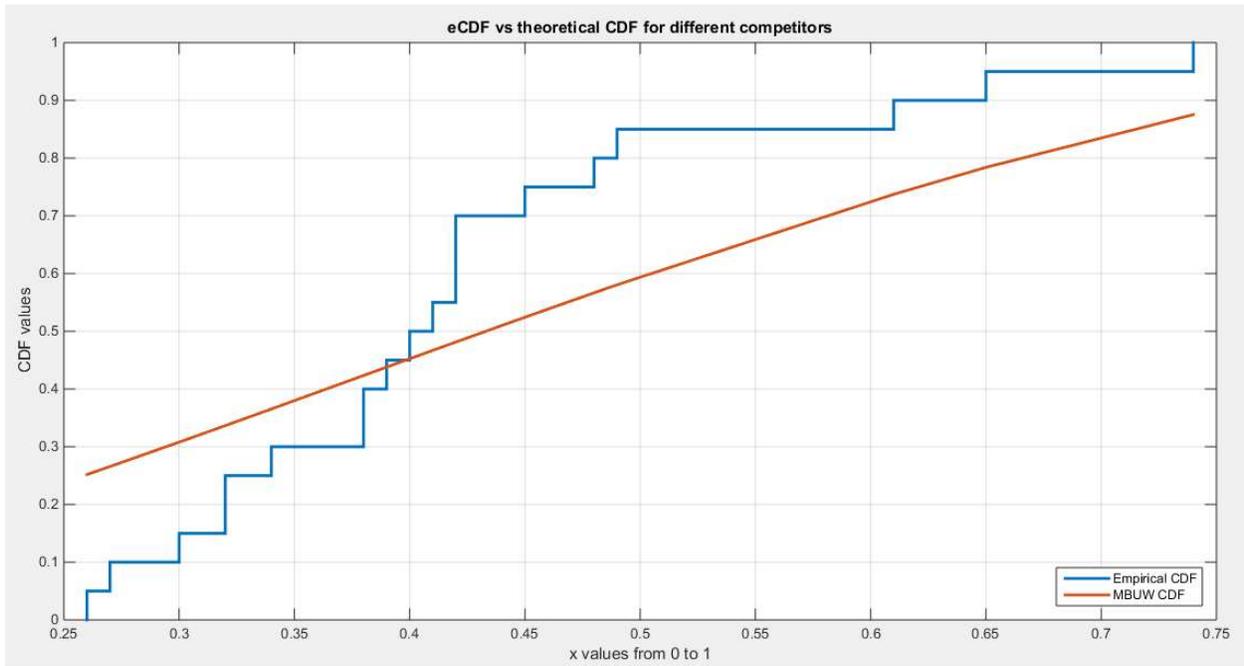

Fig. 10: shows eCDF vs. theoretical CDF of the BMUW distributions for the 4${}^{th}$ data set.

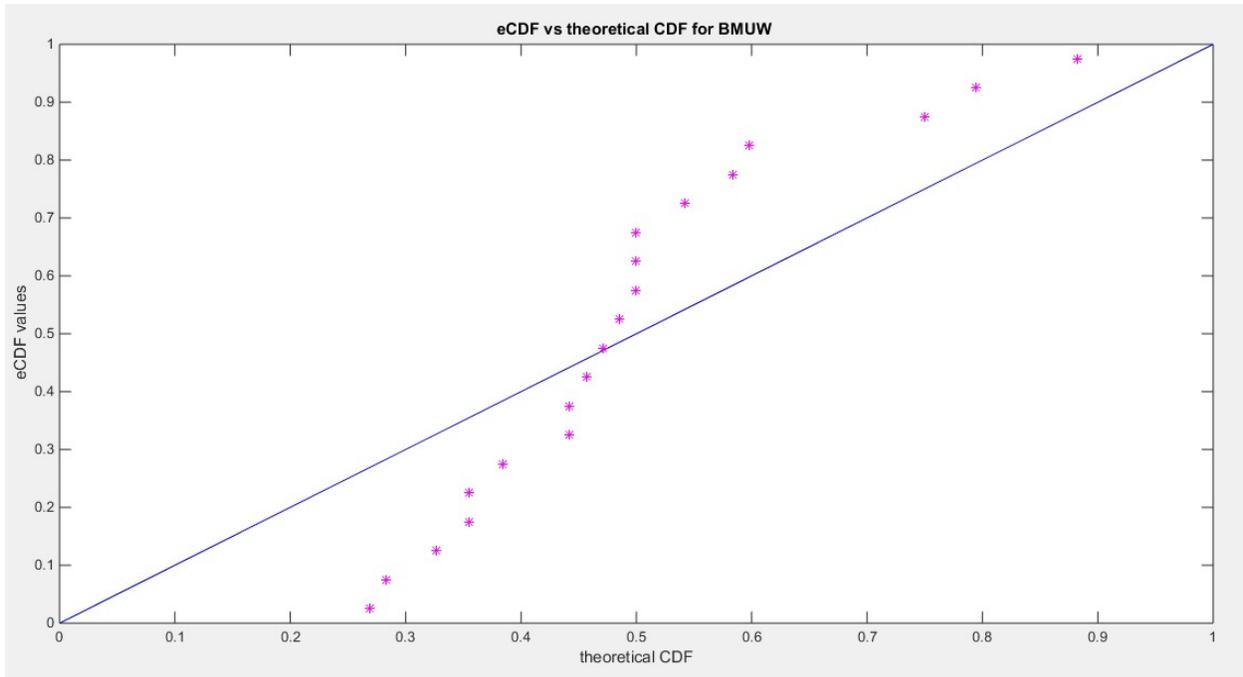

Fig. 11 :shows the pp-plot of BMUW distributions for the 4th data set.

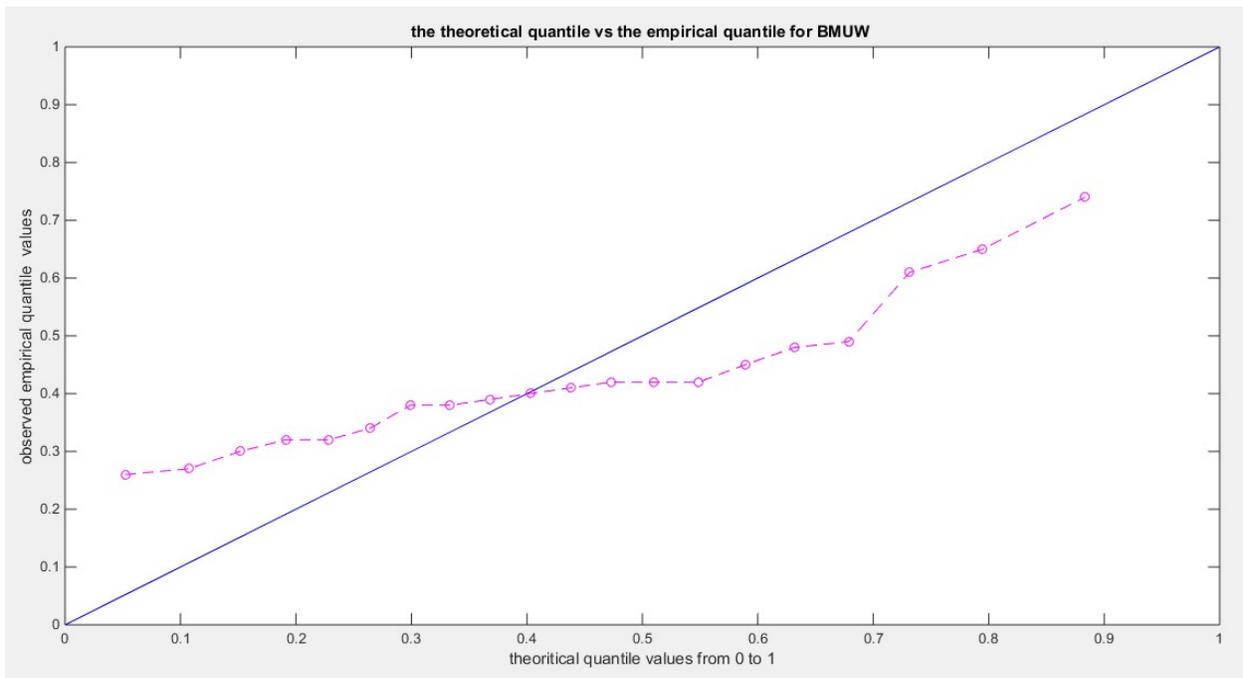

Fig. 12 :shows the qq-plot of BMUW distributions for the 4th data set.

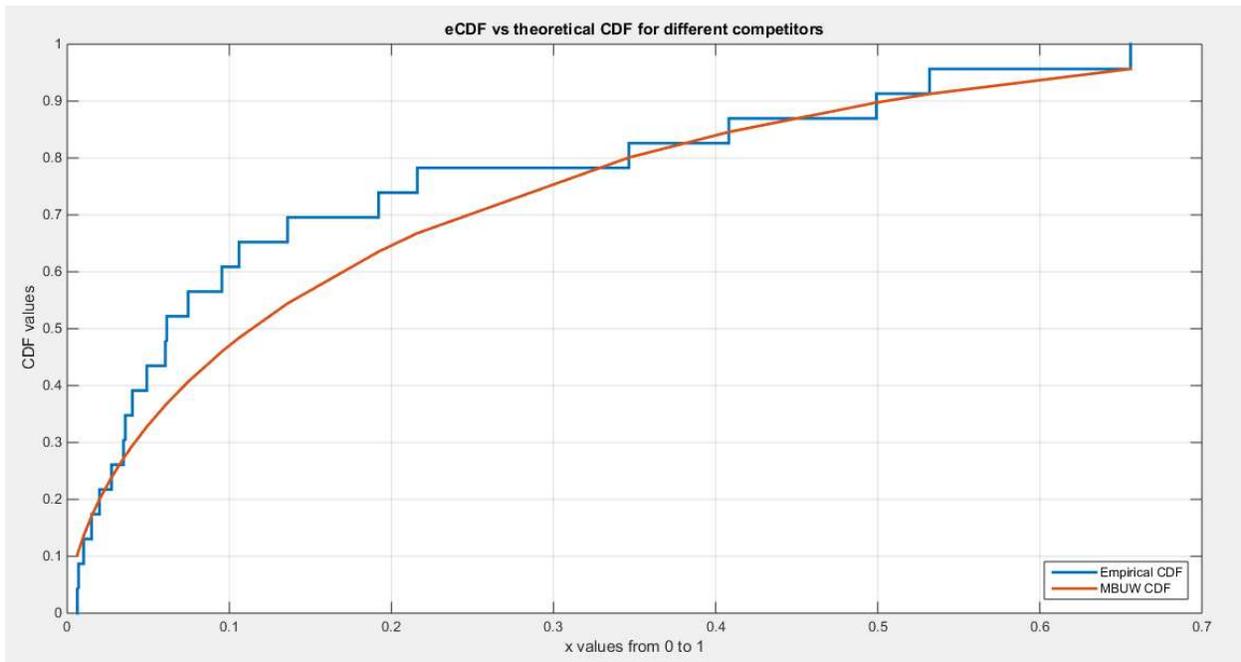

Fig. 13: shows eCDF vs. theoretical CDF of the BMUW distributions for the 5$^{th}$ data set.

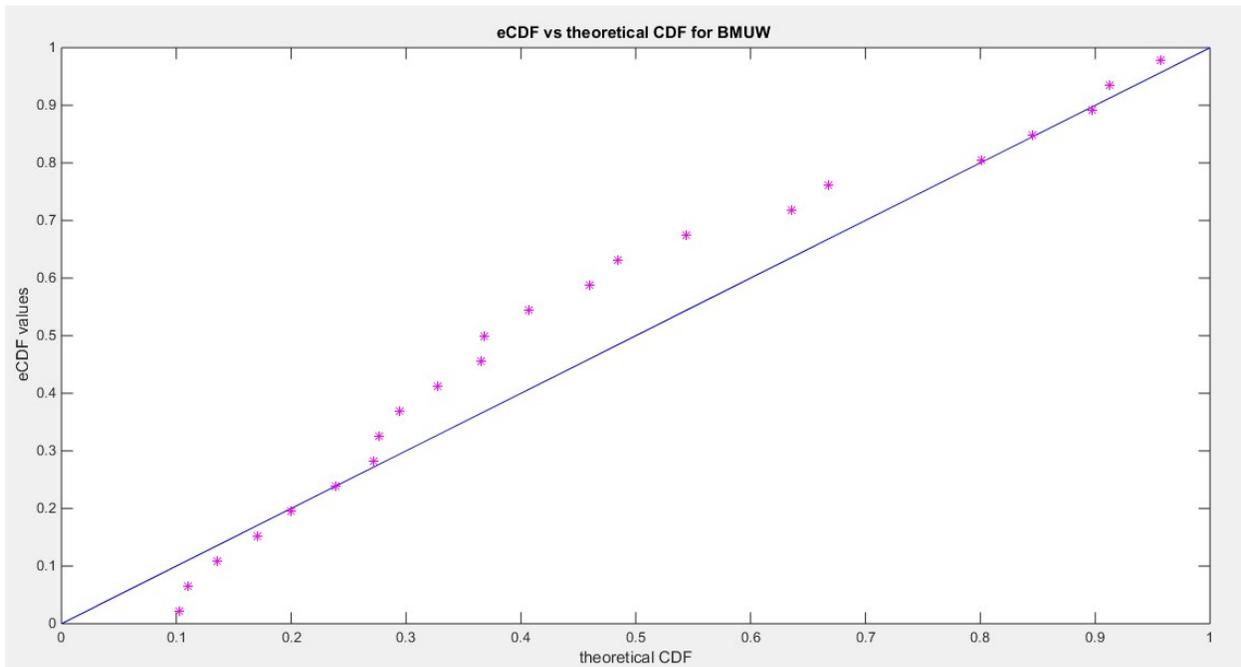

Fig. 14 :shows the pp-plot of BMUW distributions for the 5$^{th}$ data set.

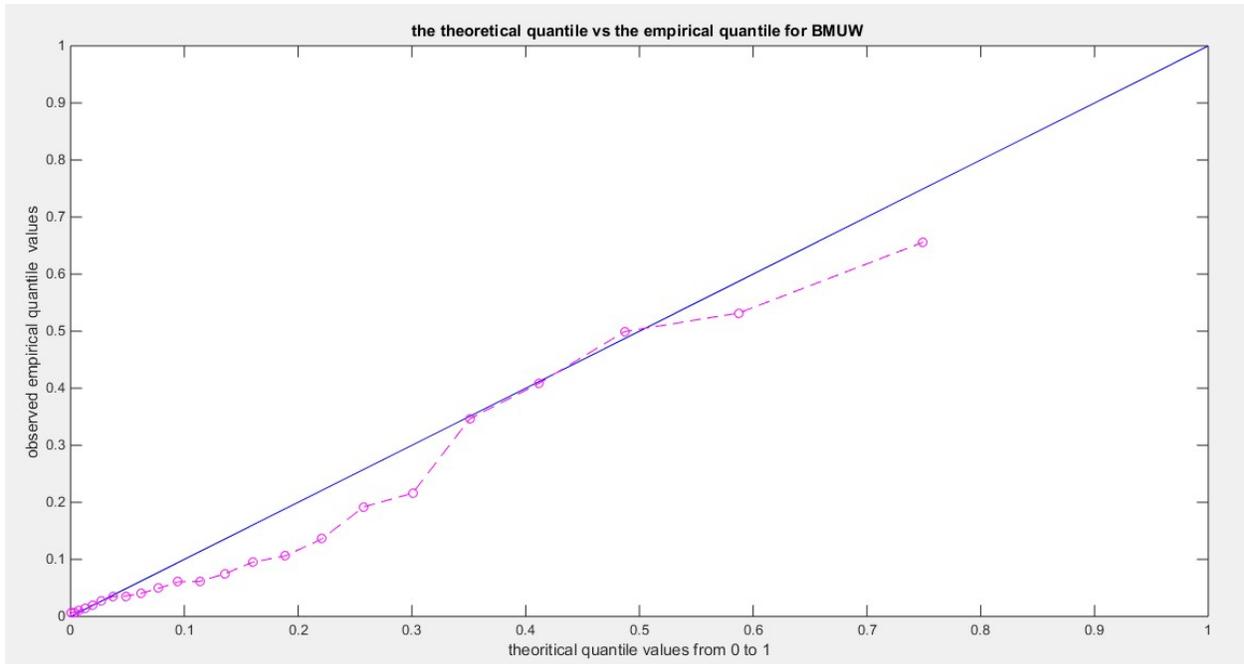

Fig. 15 :shows the qq-plot of BMUW distributions for the 5[th] data set.

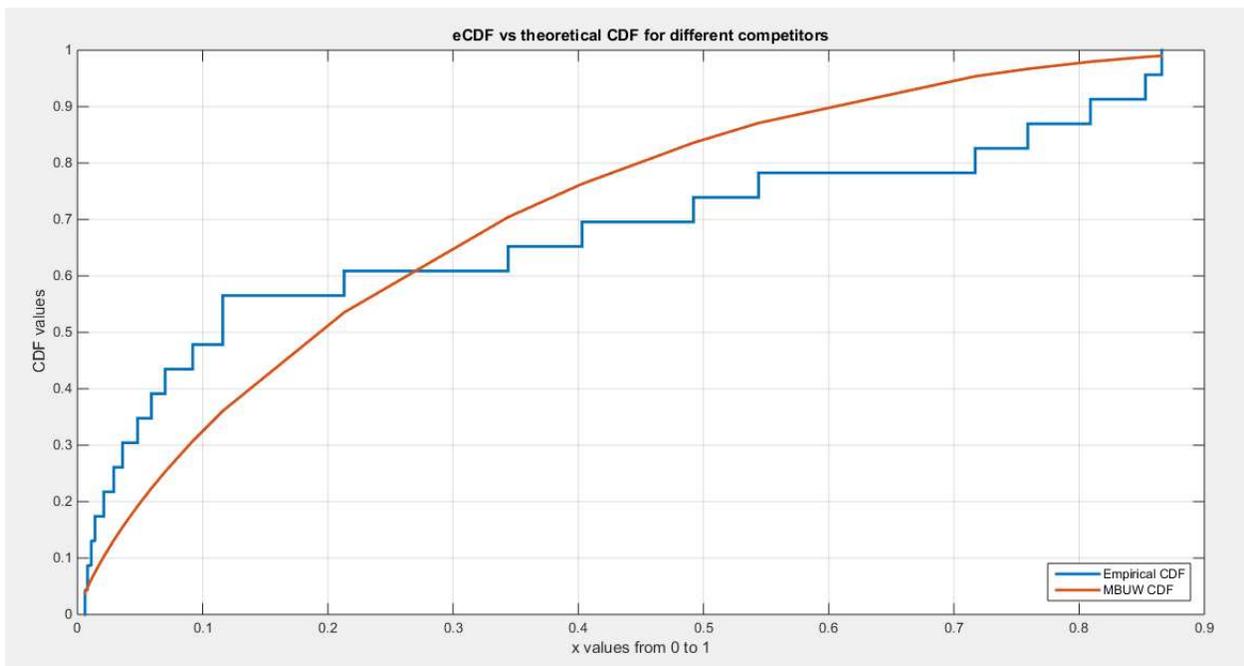

Fig. 16: shows eCDF vs. theoretical CDF of the BMUW distributions for the 6[th] data set.

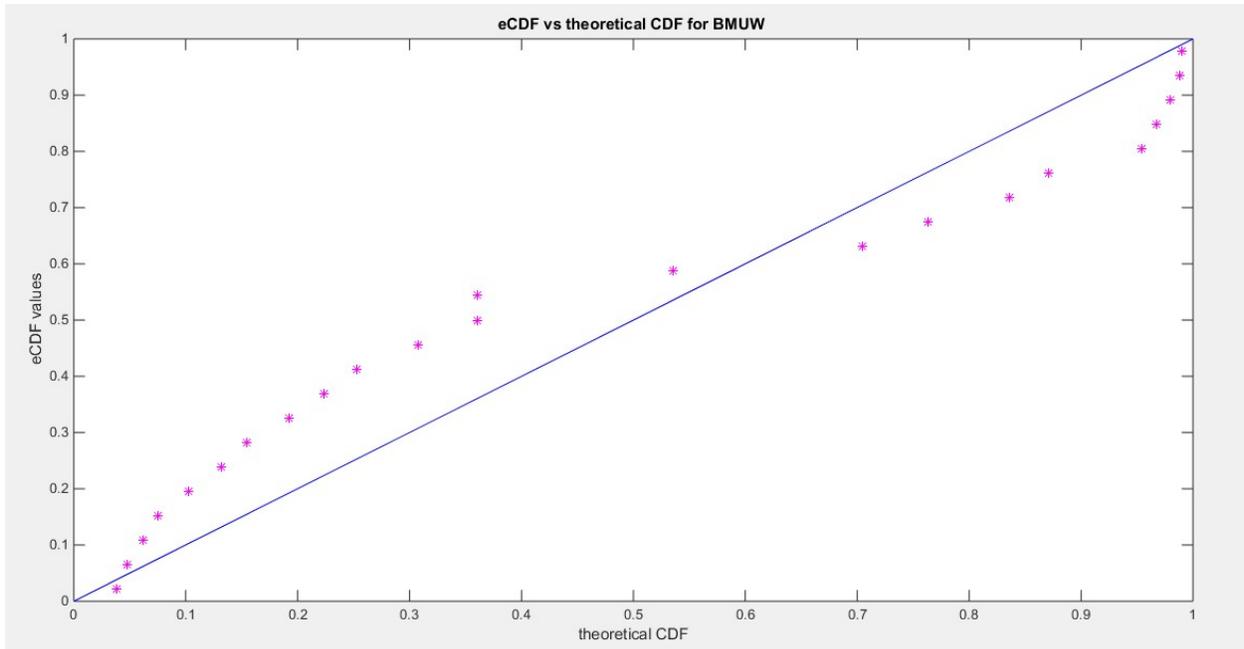

Fig. 17: shows the pp-plot of BMUW distributions for the 6[th] data set.

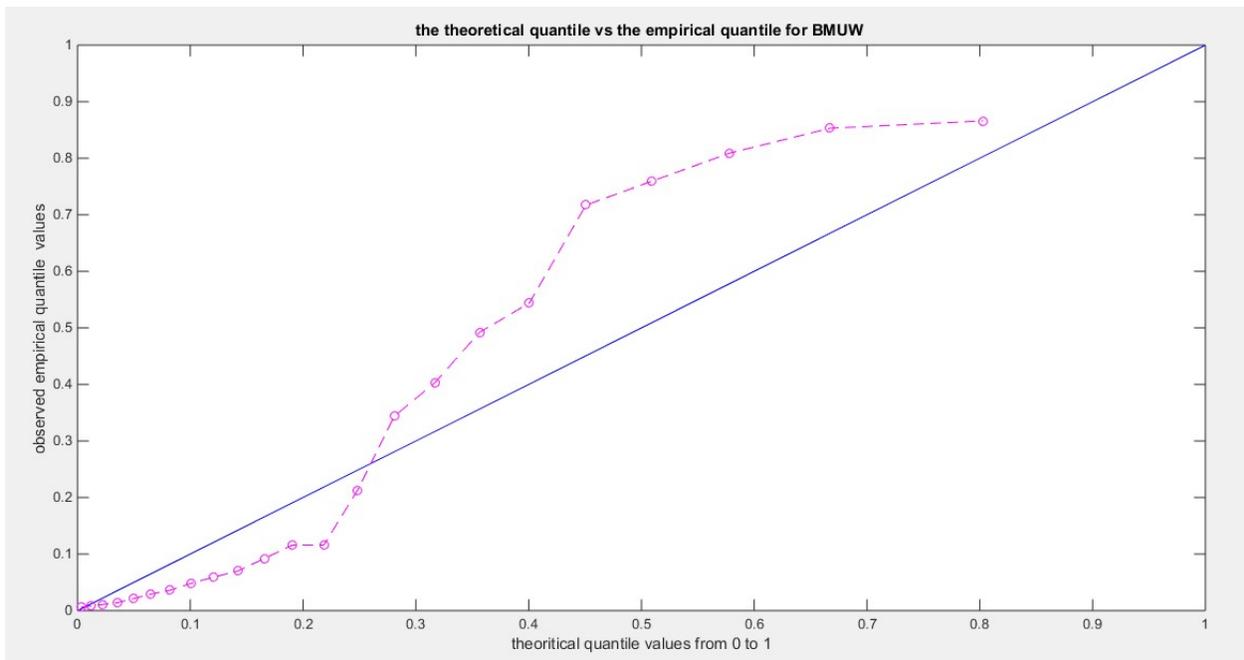

Fig.18: shows the qq-plot of BMUW distributions for the 6[th] data set.

The figures show the ecdf for the MBUW versus the theoretical MBUW, it is a more or less fit to the tested distribution as $h_0$ hypothesis cannot be failed. The sum square of error is small. The variances show dramatic reduction as compared to the values obtained from MLE estimation. The values obtained by GMMs can be used as initial guess in other modified MLE methods or other methods like least square or weighted least squares.

**Analysis of the data sets using the percentile method:**

The same datasets were used for analysis using the percentile method. The initial guess for the LM algorithm were the values obtained from MLE estimation, this worked for some data sets but it failed to give promising results as a valid KS-test ( h0= fail to reject the null hypothesis that assumes the BMUW fits the data) . The data sets that give results of fitting the distribution were the quality support and the voter turnout (The second and third data sets).

The following table shows the results of percentile method:

|  | Second data set n=20 | | Third data set, n=38 | |
|---|---|---|---|---|
| theta | $\alpha = 0.2577$ | | $\alpha = 0.6062$ | |
|  | $\beta = 1.5071$ | | $\beta = 1.2669$ | |
| Var | 52.4635 | 219.6857 | 55.7481 | 226.1499 |
|  | 219.6857 | 949.066 | 226.1499 | 945.8367 |
| SE | 1.619 | | 1.211 | |
|  | 6.889 | | 4.989 | |
| SSE | 0.0072 | | 0.0536 | |
| K-S Value | 0.1297 | | 0.1134 | |
| $H_0$ | Fail to Reject | | Fail to Reject | |
| P-value | 0.8477 | | 0.4106 | |
| $y_{25}$ | 0.865 | | 0.62 | |
| $y_{75}$ | 0.95 | | 0.78 | |

The following figures illustrate the ecdf, pp-plot and qq-plot for the results obtained from the MBUW fitting the 2 data sets using the percentile method.

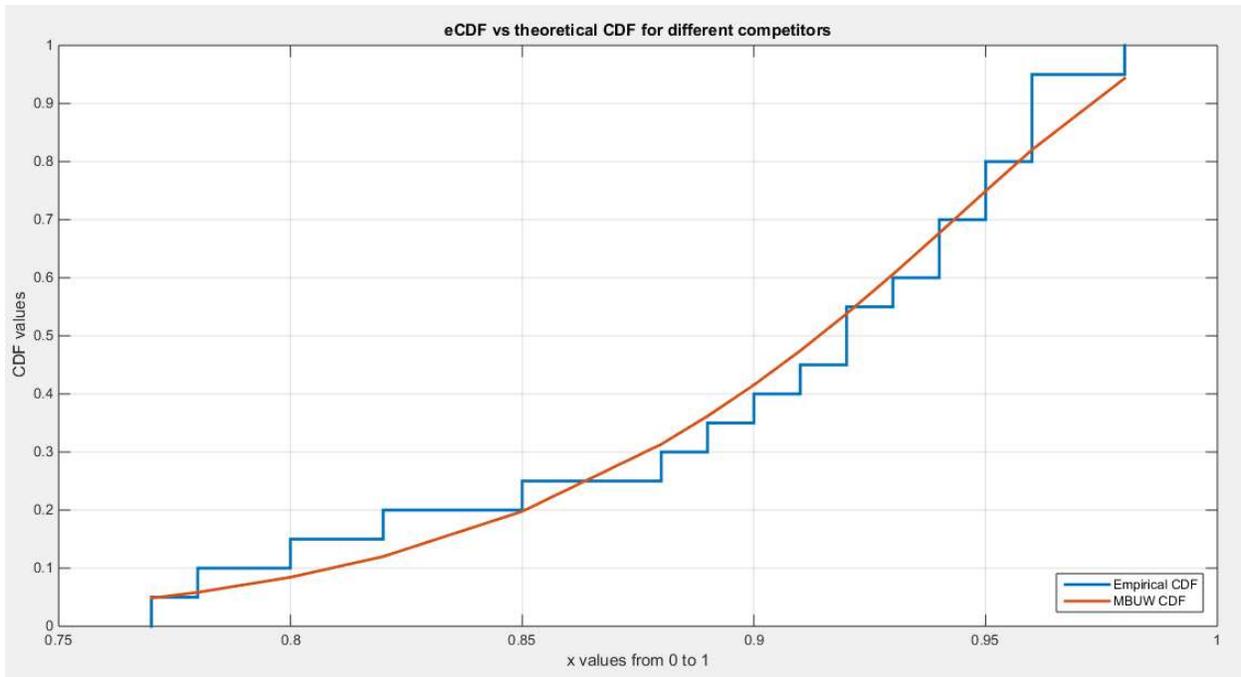

Fig. 19: shows eCDF vs. theoretical CDF of the BMUW distributions for the 2$^{nd}$ data set.

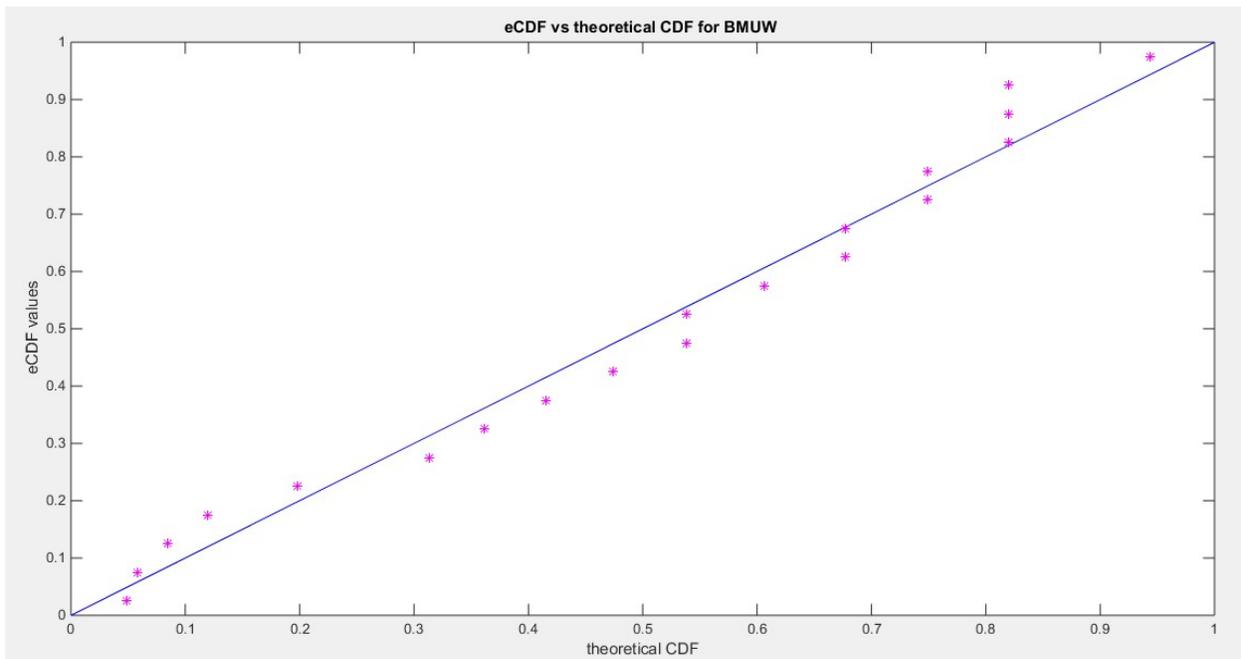

Fig. 20: shows the pp-plot of BMUW distributions for the 2$^{nd}$ data set.

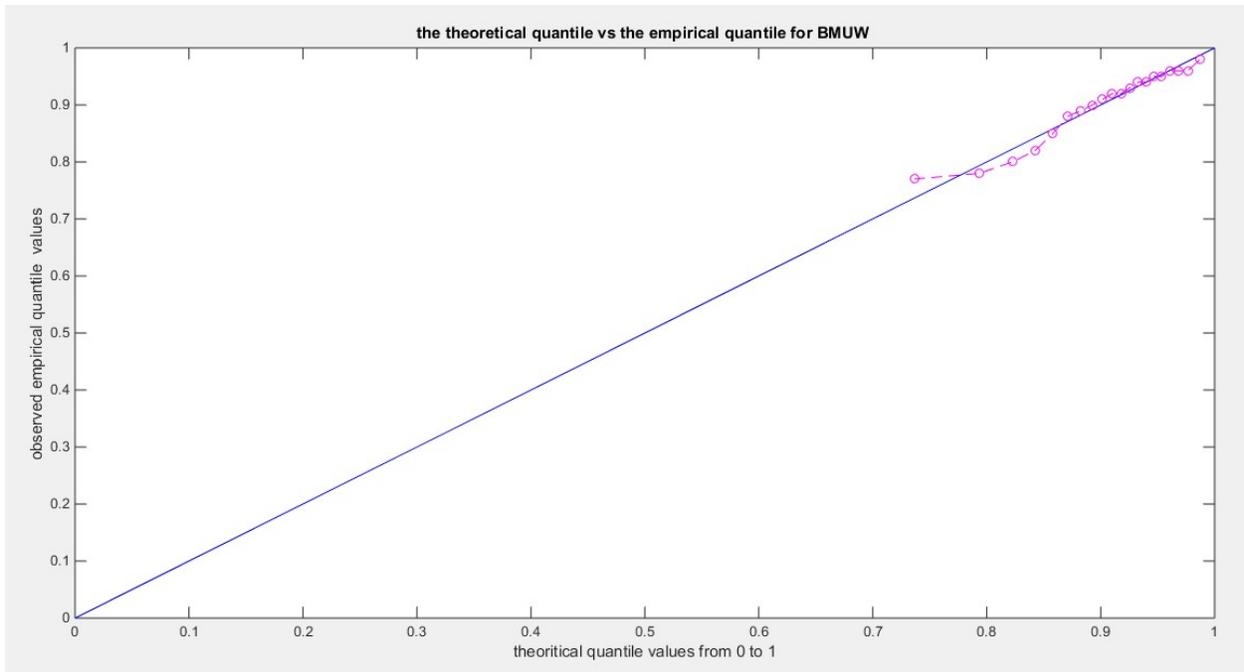

Fig. 21: shows the qq-plot of BMUW distributions for the 2$^{nd}$ data set.

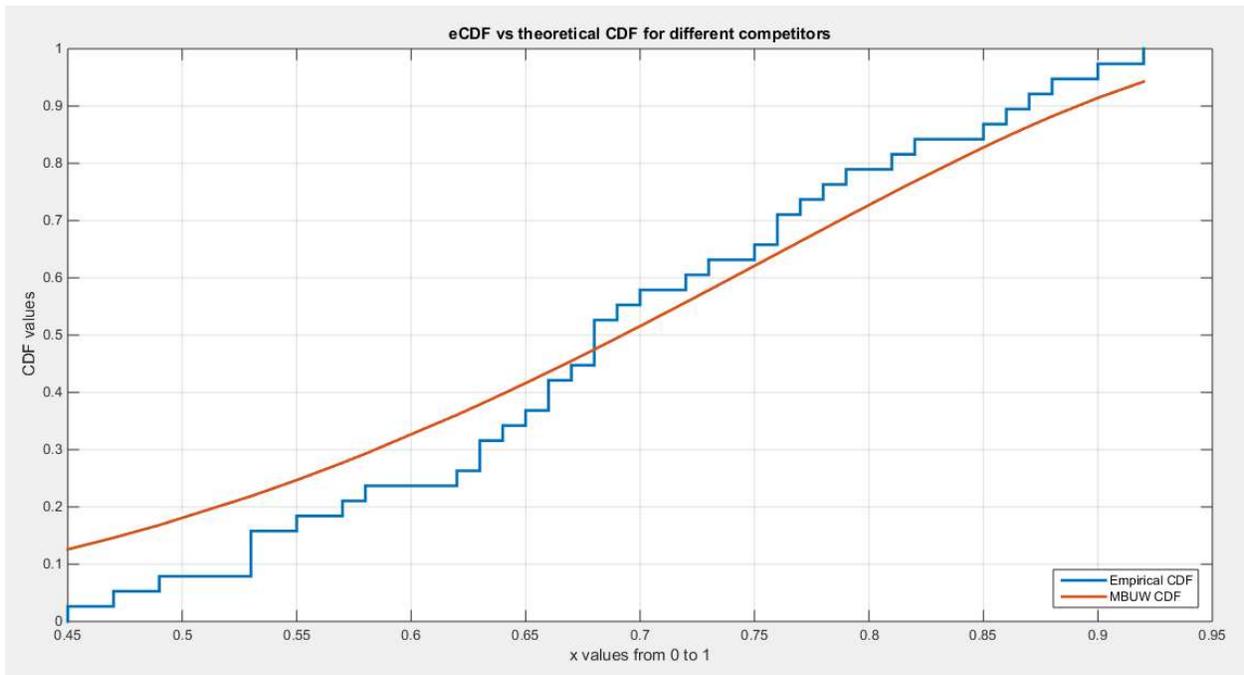

Fig. 22: shows eCDF vs. theoretical CDF of the BMUW distributions for the 3$^{rd}$ data set.

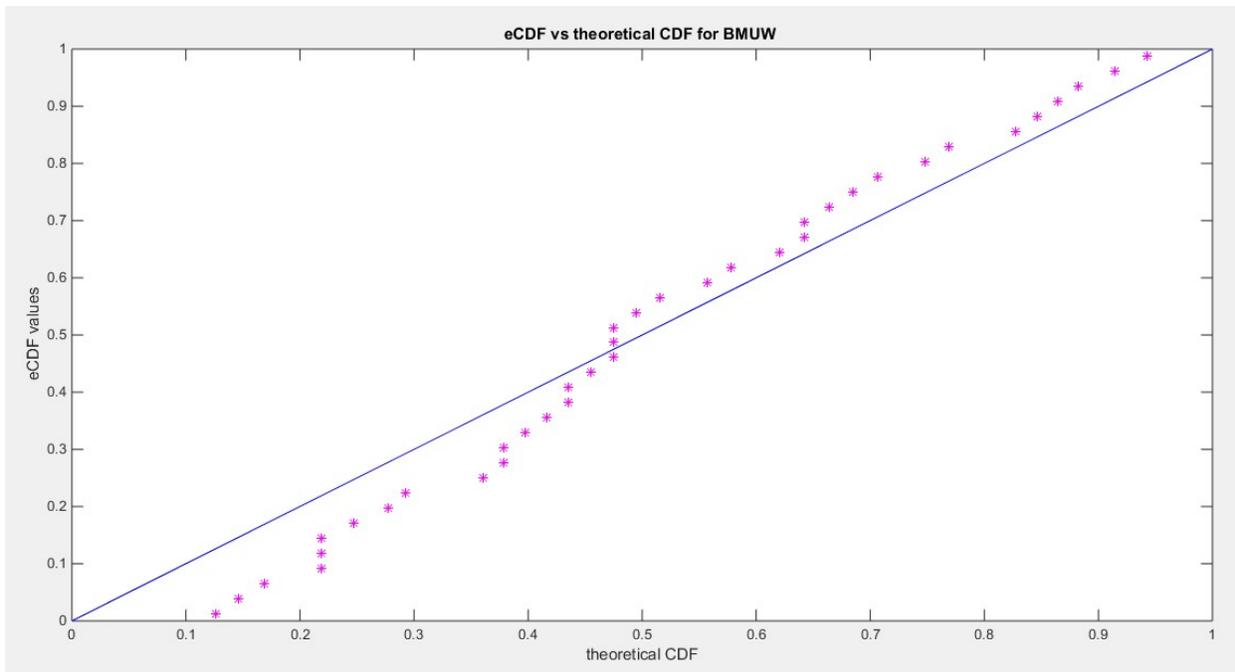

Fig. 23: shows the pp-plot of BMUW distributions for the 3$^{rd}$ data set.

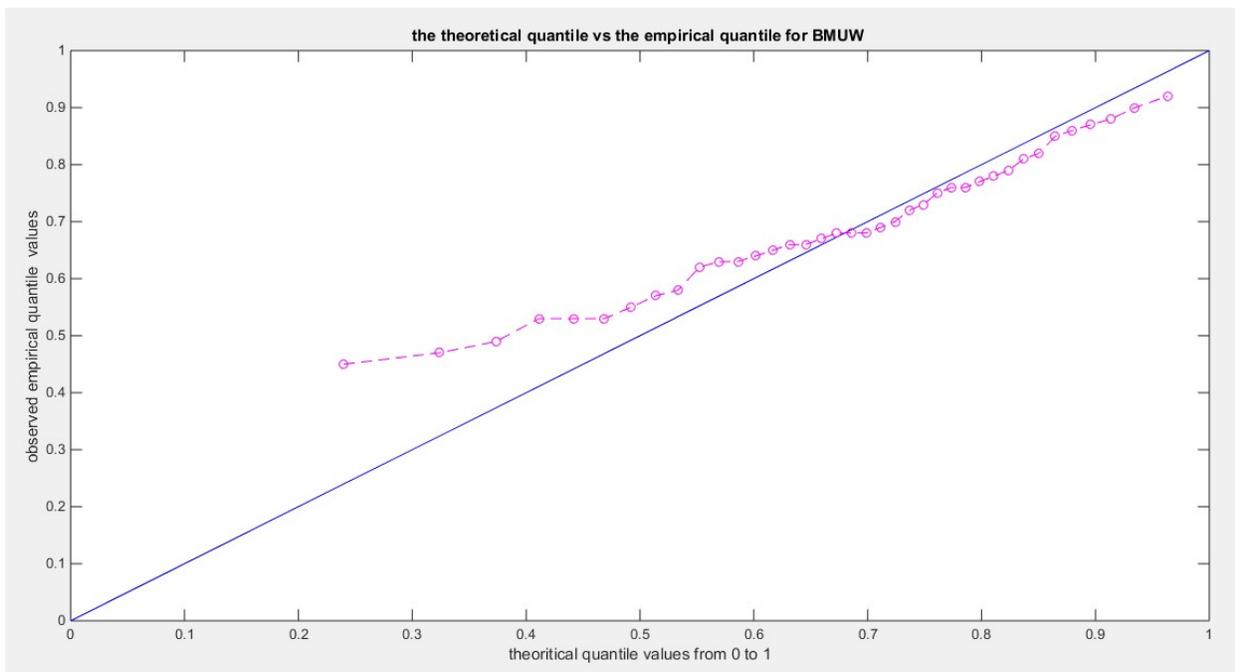

Fig. 24: shows the qq-plot of BMUW distributions for the 3$^{rd}$ data set.

**Discussion of the results:**

The analysis of the previous data sets using GMMs with the initial guess for the LM algorithm resulted in values of estimated parameters more or less equal to that given using the MLE but with dramatic reduced variance . This allows standard errors to be more accurately estimated. The percentile method was applicable to only 2 data sets. The variance obtained from this method, although it is less than that estimated from MLE, it is larger than that estimated from GMMs methods. LM algorithm is more useful in estimation as its results are more stable than other optimization algorithm. The estimated parameter values obtained from MLE method, although it may unstable with this large variance, it could be used as an initial guess in other methods like GMMs and percentile method. Moreover, the values obtained from these GMMs and percentile method can be used as an initial guess to modified MLE method and other methods like least square method (LS), weighted least square method (WLS), Anderson Darling Estimator (ADE), Cramer Von Mises Estimator (CVM), and Maximum Product of Spacing (MPS). Because the initial guess is really a challenging task, the GMMs and percentile may help in this task. The KS-test results failed to reject the null hypothesis, so it is most probably that the data follow the BMUW distribution with the p-value reported for each data set in each method. The pp-plot and qq-plot figures can visualize the fitness of the data to the distribution, but they are subjective tests rather than the KS-test which is more objective.

## Stage 4:

**Conclusion:**

This new approach of the GMMs method and percentile method can enhance the estimation process for the parameters of BMUW distribution with more stable and reliable parameter estimate and small variance. The estimators can be used as an initial guess in other methods that require numerical computations. The 2 parameters of the MBUW are difficult to be defined in term of each other so the approach described in the paper can compute both simultaneously. This is done by solving differential equation using one of the iterative techniques like Levenberg Marquardt algorithm.

**Future work:**

The author is working on modified maximum likelihood methods of estimation and other methods for this new distribution.

**Declarations:**
**Ethics approval and consent to participate**
Not applicable.
**Consent for publication**
Not applicable
**Availability of data and material**
Not applicable. Data sharing not applicable to this article as no datasets were generated or analyzed during the current study.
**Competing interests**

The author declares no competing interests of any type.
**Funding**
No funding resource. No funding roles in the design of the study and collection, analysis, and interpretation of data and in writing the manuscript are declared
**Authors' contribution**
AI carried the conceptualization by formulating the goals, aims of the research article, formal analysis by applying the statistical, mathematical and computational techniques to synthesize and analyze the hypothetical data, carried the methodology by creating the model, software programming and implementation, supervision, writing, drafting, editing, preparation, and creation of the presenting work.
**Acknowledgement**
Not applicable